# Continual Developmental Neurosimulation Using Embodied Computational Agents

Bradly Alicea[1,2], Rishabh Chakrabarty[3], Stefan Dvoretskii[4], Akshara Gopi[5], Avery Lim[1,6], and Jesse Parent[1,7]

**Abstract**

There is much to learn through synthesis of Developmental Biology, Cognitive Science and Computational Modeling. Our path forward involves a design for developmentally-inspired learning agents based on Braitenberg Vehicles. Continual developmental neurosimulation allows us to consider the role of developmental trajectories in bridging the related phenomena of nervous system morphogenesis, developmental learning, and plasticity. Being closely tied to continual learning, our approach is tightly integrated with developmental embodiment, and can be implemented using a type of agent called developmental Braitenberg Vehicles (dBVs). dBVs begin their lives as a set of undefined structures that transform into agent-based systems including a body, sensors, effectors, and nervous system. This phenotype is characterized in terms of developmental timing: with distinct morphogenetic, critical, and acquisition (developmental learning) periods. We further propose that network morphogenesis can be accomplished using a genetic algorithmic approach, while developmental learning can be implemented using a number of computational methodologies. This approach provides a framework for adaptive agent behavior that might result from a developmental approach: namely by exploiting critical periods or growth and acquisition, an explicitly embodied network architecture, and a distinction between the assembly of neuronal networks and active learning on these networks. In conclusion, we will consider agent learning and development at different timescales, from very short (<100ms) intervals to long-term evolution. The development, evolution, and learning in an embodied agent-based approach is key to an integrative view of biologically-inspired intelligence.

[1] Orthogonal Research and Education Laboratory, Champaign-Urbana IL USA
bradly.alicea@outlook.com
[2] OpenWorm Foundation, Boston MA USA
[3] Independent Researcher, Paris FRANCE
[4] Technical University of Munich, Munich GERMANY
[5] Manipal Institute of Technology, Manipal INDIA
[6] Plot Twisters, Virtual (https://www.plottwisters.org/)

## 1 Introduction

The process of biological development provides many novel lessons for embodied approaches to computational agents. Development serves as an inspiration for artificially intelligent processes such as the serial acquisition of linguistic information [1]. This can be summarized as a compositional approach to development. While serial acquisition can capture the linguistic and psychological aspects of information acquisition in a naively intelligent system, it is largely insufficient to properly characterize the developmental process that creates an embodied nervous system. Even when developmental processes are made explicit in the algorithm [2], they often focus on the psychological (particularly constructivist) nature of development [3, 4]. One alternative involves the exploration of developmental transformations in growth and connectivity, particularly as they relate to an agent's body and neuronal network topology [5]. A neuronal network is different from a conventional neural network in that sensory input results in motor output with both linear and nonlinear components. This connectivity matrix depends on edges connecting the network's individual processing units, and so can change both in terms of size and reconfigurability over the course of development. In short, dBV neuronal networks are analogous to small animal connectomes, where small circuits can regulate simple behaviors. Elman pointed out in 1993 [6] that in the context of learning language, the connectionist approach fails except in cases where the network is subject to a growth process where the topology starts small then expands. Yet even in this case, such studies do not explicitly leverage embodiment as a fundamental aspect of the developmental process. Here, we use developmental Braitenberg Vehicles (dBVs) [7] to understand the developmental process via two different frames of analysis: 1) the level of network topology and 2) the impact of both innate or genetically-defined factors (the initial condition for any phenotypic or behavioral trait) and external environment on agent phenotype. Genetically-defined factors are indeed directly influenced by the external environment through plasticity. A plastic response provides the capacity of an agent to change or exhibit variation among its parts when responding to an environmental stimulus.

Braitenberg Vehicles [8] serve as toy models of an embodied nervous system with a simple input/output relationship between sensors and effectors. In a Braitenberg Vehicle, the relationship between sensor and effector is a white box relationship [9]. A minimal set of connections are made, and conduct a sensorimotor signal that can either be linearly mapped from sensory to effector or experience contralateral cross-talk. This occurs when sensors from one side of the vehicle carry a signal to an effector on the opposite side. dBVs (see Figure 1) refine this basic model in three ways: a generative model of a multicellular agent with defined intercellular communication channels, an increase in representational complexity over developmental time, and allow for multiple hidden layers between sensor and effector [10]. These two points can be collectively understood as connectogenesis [11] in the biological context, and refers to the generativity and emergence of this process. Indeed, these properties often capture the flexible and sub-optimal wiring patterns in biological brains [12]. Thirdly, learning is enabled by different types of acquisitions called morphogenetic (structural) and behavioral (functional), which themselves are further enabled by a substrate determined by innate capabilities. Developmental neurosimulation offers a balanced approach to learning: growth of the internal network, plasticity from an initial set of innately-defined conditions, and morphogenesis (the process of growth and transformation of the agent's neuronal network and body) of the agent's body all enable a naturalistic agent containing a developing morphology and plastic neuronal network.

We will utilize the dBV approach to explore the nature of developmental learning and the heterogeneous timing of information acquisition. The first aspect of this can be summarized as *developmental freedom*, which summarizes the effect of network reconfigurations given connectivity constraints of the neuronal network and plasticity at the network level. A system exhibiting developmental freedom demonstrates the ability to produce alternative network configurations given connectivity constraints of the neuronal network and plasticity at the network level. A greater amount of freedom results from an increased amount of reconfigurability and ability to produce nonlinear outputs, with spatial embodiment [13] providing context for learning. In Continual Developmental Neurosimulation, *spatial embodiment* is a developmental antecedent of learning, but also provides a substrate for learning. Alternate arrangements of sensors, effectors, neuronal network connectivity, and phenotypic geometry results in different outcomes from similar sets of innate capabilities and environmental experiences. The third aspect is better known as the *critical period* of development, or an enhanced period of learning enabled by a heightened nervous system plasticity. While the critical period itself is determined through genetic mechanisms, its effect on learning is environmentally variable. Depending on when the critical period occurs relative to sensory acquisition, the agent can learn a vast amount or miss out on learning altogether. Collectively, developmental freedom, spatial embodiment, and critical period regulation contribute to a form of artificial learning with explicit developmental mechanisms.

**1.1 Deep Learning Approaches to Continual Learning.** Continual learning has been explored in the world of deep learning in ways that are distinct from our approach. In deep learning-related bio-inspired approaches, the focus is on implementing continual learning as a means to avoid catastrophic forgetting. Lassig et.al [14] provide three main strategies for implementing continual learning in a computational agent: methods for reviewing data from old tasks while engaging in new tasks (replay), constraining current learning in order to protect previous learning (regularization), and selective malleability of weights and neurons in the network (architecture). For deep learning approaches, the focus is on learning specific tasks, which are the impetus of adaptation, memory maintenance, and generalization. In dBVs, acquisition via embodied context (shape and relative size of body parts) for specific tasks are connected to potential task-independent behavioral responses [15].

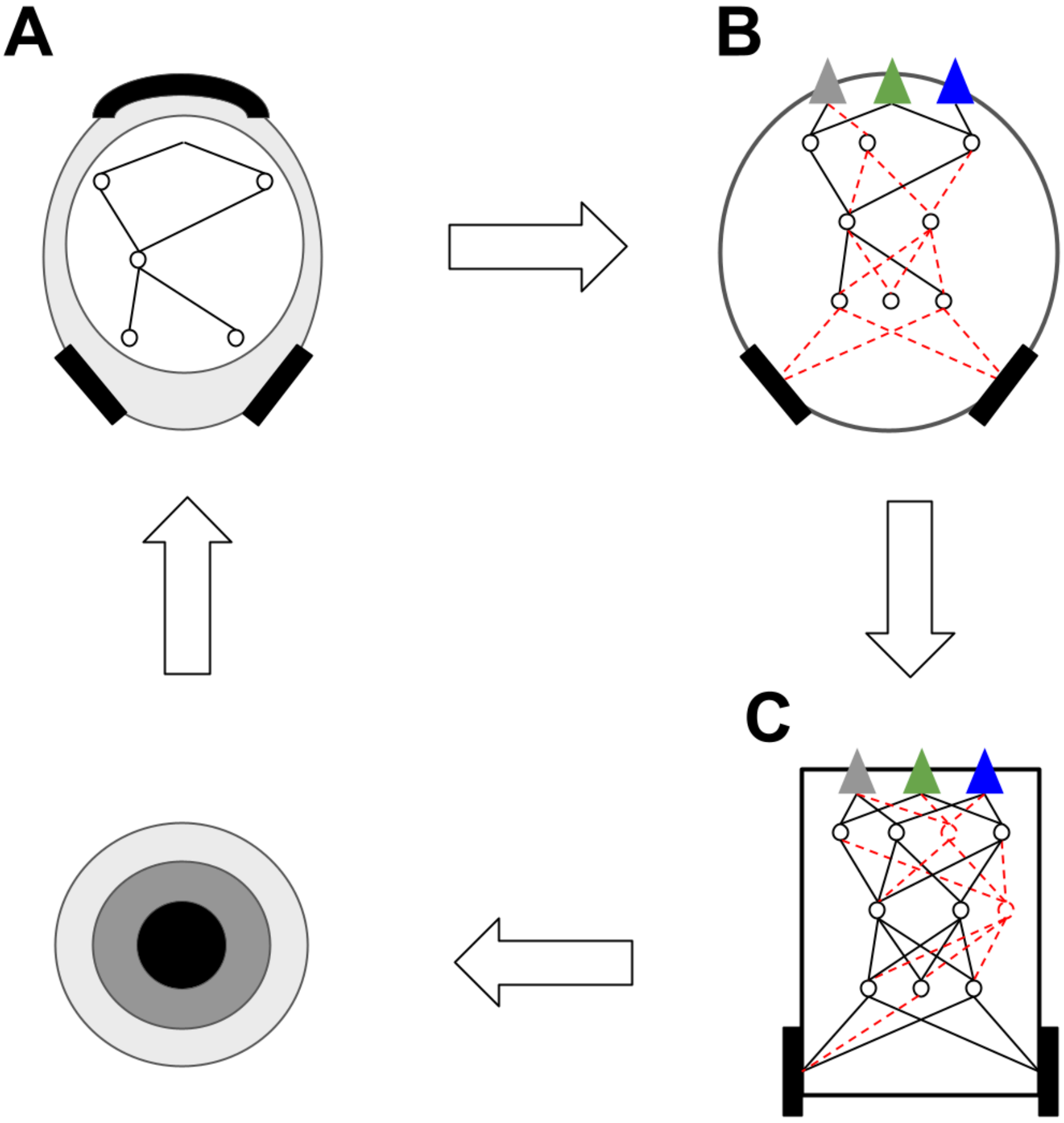

Figure 1. Three stages of dBV neuronal network development. Examples A, B, and C occur sequentially in time. Example C can also reproduce, giving rise to the three layered starting form at the lower left. Morphogenetic acquisitions during this example are strictly cumulative: nodes and connections are added but never lost, which is not the case in typical dBV implementation. A: a dBV with three light sensors (gray, green, and blue channels) connected to an internal network with a small set of nodes and connections (black circles and lines, respectively). The neuronal network output drives two effectors (left and right wheels). B: a dBV where example A's features are retained, but with the birth of additional nodes and connections (red dotted circles and lines, respectively). C: a dBV where features of example B are retained, but with the birth of additional nodes shown in example B.

Replay and regularization rely on implementing a form of the stability-plasticity tradeoff [16], or the need to maintain stable memories while also enabling adaptive changes in synaptic weights. The stability-plasticity tradeoff also plays a role in age-related learning constraints [17], which we will revisit later in this paper. This suggests a generalized homeostatic mechanism, which is responsible for behavioral robustness and associated resiliency [18]. The ability to capture adequate context, particularly in the form of multisensory integration [15, 19], is key to enabling agent generalization. Implementing a suitable architecture relies on modeling plasticity and neurogenesis [20], which can enable reconfiguration and repair.

Critically, conventional deep learning methods do not focus directly on input-output relations. This is in line with a focus on discrete tasks: the connection between inputs shaping an internal representation and a wide range of output behaviors over different stages of an agent's interactions are irrelevant. Task structure is of primary importance: temporal structure and relation to other tasks impacts performance [21]. Of more immediate interest is the replay of specific experiences, which work on the level of dynamic stability and recall ability [22]. More important to continual learning is the encoding of task-specific architectural features that generalize across multiple tasks [23]. To do this, methods such as Deep Feedback Control [14] and Synaptic Intelligence [24] are implemented to maintain robustness to task learning of various types.

In [15], the authors propose several biological mechanisms that support lifelong learning in computational agents. We will focus on the following: metaplasticity, cognition outside the brain, reconfigurable organisms, and multisensory integration. We also present additional mechanisms that are important to enabling continual cognition and behavior, chief among them a connection between reconfigurability and embodiment. The developmental neurosimulation approach also offers an embodied dimension to this literature, which provides new challenges and opportunities.

### 1.2 Development, Learning, and Developmental Freedom

In dBVs, development and learning are nominally independent: development is an anatomical phenomenon, while learning is a representational one. Taken collectively, the process of a dynamic neuronal network defines the scope of developmental freedom, or the degrees of freedom enabled by the architecture. Developmental freedom is the tendency for information and representations in particular to take advantage of a stable underlying topology. This can be understood in relation to contingency. Developmental contingency restricts the developmental neuronal network to an increasingly limited set of possible trajectories. Contingency results in earlier developmental events (cell birth or the establishment of connections) to constrain the range of possibilities for future developmental events (formation of a network motif). Furthermore, contingency locks the neuronal network into an increasingly smaller number of pathways over developmental time, and results in path dependency. Once the network configuration becomes formalized, formation of a stable but robust circuit benefits from developmental freedom.

In Figures 1 and 4, we introduce an embryogenetic model of Braitenberg vehicle development. The standard Braitenberg vehicle is shown to develop from an undifferentiated initial state (analogous to an egg), denoted by a single spherical object. The next stage of development involves either diploblastic (two-layered) or triploblastic (three-layered) concentric layering. Layered structures (blastic structures) establish the basis for various functional components of the agent body. For the three-layered structure shown in Figure 1, the inner, middle, and outer layer become raw material for sensors/effectors, body shell, and the neuronal network chamber, respectively. As the dBV matures, these layers separate and migrate to their adult locations. For example, a single layer (sensors/effectors shown in black on Figure 1) can split into four parts, two migrating towards the vehicle front (sensors) and two migrating towards the back (effectors). An agent that starts from an undifferentiated initial state with three layers experiences morphogenesis to reach its mature form as a Braitenberg vehicle. Components for each layer allow for compositional phenotypes, which can be mutated and recombined during development. This can be determined via the genomic encoding, shown in Figure 2.

To discuss developmental freedom in the context of Braitenberg Vehicle nervous systems, we must first consider how nervous systems connect together during development. The time window model of Lim and Kaiser [25] suggests that this process is a highly parallel process where the state of connectivity at any one developmental time point can have far-reaching effects across an agent's life-history. A model of contingency (see Figure 1) allows us to think about the mapping between sensor and effector in terms of branching developmental stages, while each time point serves to determine the trajectory of future growth and development. Recent work [26] has also brought into question that neuronal network assembly occurs exclusively via shortest paths, resulting in a sprawling constellation of alternative network communication models. In developing nervous systems, complexity accumulates through the duplication and divergence model [27, 28], wherein mappings between sensor and effector can be reused and duplicated as sensors, effectors, and nervous system modules are duplicated and deployed to yield more complex phenotypes. This expansion of connectivity between sensor and effector provides a more flexible network than a standard neuronal network, and is enabled by spatially-explicit maps between environmental features and topology in the internal network. We propose that the combination of connectogenesis, artificial morphogenesis, and direct sensory processing enable neurosimulation in ways that simply taking weights and nodes in layers of a disembodied network cannot.

*1.2.1 Developmental Freedom as an Emergent Phenomenon.* As the neuronal network is expressed, it serves as a substrate for information processing similar to that of morphogenetic and behavioral substrates at the periphery. However, rather than being exposed to environmental information or serving as the scaffolding for other traits, the neuronal network serves as a substrate for information transfer between the sensor and effector. There are a number of pathways from sensor to effector, which grow exponentially as the neuronal network grows in size. To this end, growth and transformation of the phenotype also contribute to developmental freedom, particularly in terms of where and how sensory inputs are established. Changes in sensory input, spatial arrangement of the agent phenotype, and phenotypic symmetry can also affect the underlying substrate. Developmental freedom is the tendency of information processing to take new pathways along the substrate. While not autonomous itself, developmental freedom is advantageous, and can be related to ontogenetic strategies often observed during development [11, 29]. This ability to find new paths comes from the stability of a network that is no longer growing. For example, exploratory behaviors cannot stably emerge on networks that are constantly reconfiguring. The best and shortest pathways are constantly changing, undercutting the efficiency of information processing. Similarly, during connectogenesis, the potential for pruning connections due to lost inputs or the emergence of feedback loops can fundamentally alter information processing itself. Thus freedom on the substrate enables learning and stable behaviors to emerge.

### 1.3 Spatially-dependent Embodiment

There are a number of interesting connections between dBVs and their embodiment. The first is how embodiment allows the agent to sample the incoming data in a stratified fashion: the relative position and orientation of sensors allows the agent to compare points in a differential manner. Differential comparison results from spatialized sensor arrays, which is analogous to biological sensory organs. This allows for different frames of reference to be distinguished, and is distinct from the input layer of a neuronal network. Each sensor observes the world at a specific angular perspective relative to the vehicle's midline. This constant interaction with an environment of sensory cues allows the agent to keep concepts about the world continuously up to date, and provides a contextual fabric for the internal representation [10]. Variation in bilateral symmetry may yield some interesting effects. One effect is that complex movement outputs can be amplified and modulated. Asymmetrical sensory systems can also be advantageous in learning more about the vehicle's environment. A dBVs overall bodily structure allows it to extract more information from a single data point than in a disembodied neural network. Changes in the relationship between sensor and effector provides context for simple behaviors.

### 1.4 Network Reconfiguration and the Critical Period

Another attribute of development is the unfolding of time-dependent processes. One powerful developmental principle we can draw from is the existence of a critical period [30-32]. The critical period allows for enhanced sensitivity at specific points in developmental time [33]: acquisition of sensory information and phenotypic plasticity during this time have an outsized effect on what our agent looks like and how it behaves. The critical period is also an important time for learning from sensory inputs, and loss of function during this time can radically reconfigure the neural architecture [34, 35]. In the development of touch, delays in development as compared to the normal developmental trajectory can lead to a loss of sensory processing ability [36].

Hensch [30] proposes two potential mechanisms that define enhanced neural reorganization during the developmental critical period in support of learning. First is a functional competition between inputs that might be thought of as developmental allocation. Second is a role for electrical activity in the structural consolidation of selected pathways. This has a number of consequences on the later stages of development and ultimately the adult phenotype [37]. Learning and anatomical change also seem to occur at different time scales [38], but learning in the service of behavior might be dependent on the length of time afforded to developmental growth [39].

According to this view, while both development and learning are mechanisms that induce neural and behavioral plasticity, development is a generative phenotype largely influenced by experience-expectant mechanisms [40]. As the dBV interacts with its environment, these interactions shape the manner in which the network is generated. Learning that occurs independent of large-scale changes in the network is the product of experience-dependent influence. Changes in the timing of this transition from a neuronal network largely influenced by experience-expectancy to one primarily influenced by experience-dependency is key to understanding how developmental processes shape information acquisition and the supervision of learning.

## 2 Methodological Perspectives

### 2.1 Artificial Genetics and Environment

Development is captured using a generative process of constructing nervous system nodes and weighted connections, while learning is defined by a process of connectionist plasticity. In this paper, we have instantiated these mechanisms using Genetic Algorithms and Hebbian Learning mechanisms [10] on top of an architecture for developmental innate capabilities. The generalized influence of sensory information is captured using the concept of Gibsonian Information [41]. In the proposed architecture for innateness, a genetic algorithm is used to demonstrate the link between encodings representing innate capabilities and expressed substrates for various acquisitions.

*2.1.1 BraGenBrain encoding.* The first version of the genetic encoding is utilized by the BraGenBrain implementation demonstrated in [7]. Sensors (inputs) and effectors (outputs) are connected by an artificial neuronal network, the components of which are generated by an algorithm. In an object-oriented framework, vehicles are a class, instances of which can be generated by a factory function. All generated neuronal networks are evaluated by a fitness function, characterized by force (non-inertial movement), rotation (movement trajectory curvature), and displacement (sliding movement). These also combine into a speed vector, which are linear components of a behavioral signal. Additional parameters in the BraGenBrain approach are effect strength and a specific gravity constant. BraGenBrain uses four genetic operators to determine various aspects of evolutionary change: initial population size (default value of 10), world size (with a default size of 800x800 pixels), a mutation rate (default frequency of 0.05), and crossover (random reproduction with a default frequency of 0.05). Phenotypes in new generations of vehicles are biased by natural selection (fitness) or neutral drift.

Each developmental neuronal network is represented by a directed acyclic graph (DAG) with a unique ordering. Nodes are added at a rate determined by the genomic encoding, particularly innate timing mechanisms (specified for each agent lineage, see Figure 6). Each neuron's output is normalized as synaptic weights (a total of 255), which sum up to one. A matrix graph representation is used to store connectivity information. The neuronal network is ordered from input (root) to output (tip). Root nodes of the neuronal network (proximal to the sensors) correspond to sensory inputs. Leaves of the network (in the direction of the effectors) correspond to internal movement circuits and their associated representations. Tip nodes (proximal to the effectors) correspond to movement outputs. The neuronal network architecture is structured by a developmentally-inspired three-step evolutionary algorithm. This is distinct from canalization function-driven morphogenesis, and occurs within the inner blastic layer (neuronal network chamber) among newly-born agents. Initially, a mutation occurs in the standard binary encoding at a specified probability. Vehicles are then evaluated for the next generation using the past generation's speed vectors as a criterion. Crossover allows for variation in the architecture: neuronal network components are inherited from two randomly-selected agents. This produces a hybrid offspring with a nominally high fitness.

*2.1.2 Critical Period encoding.* The second type of genetic encoding is specific to dBV agents utilizing critical periods to enable development [42]. The onset of a critical period is triggered by a binary switch, leading to alternate expressions of the plasticity encoding. This mimics a regulatory switch that can be turned on and off based on regulatory control mechanisms. Each agent also has a genetic encoding of sensory organ acuity consisting of discrete register entries representing sensor-specific acquisition (with the potential for developing differential sensory acuity) during the critical period. Direct sensory transformations provide proto-representations that extract different components of information (e.g. low-pass and high-pass). These transformations can be captured by both simple logic gates as well as more complicated functions related to reversible logic gates and canalization processes. Canalization refers to a specific developmental trajectory resulting from a series of differentiation events. Differentiation allows a layer component to acquire a more specific form or function. For example, a germ layer can divide into two spatially discrete components that later go on to produce four sensors and two effectors. Differentiation events occur in a serial fashion, with changes being acquired one at a time as shown in Figure 6. The process of canalization then is analogous to a system of channels in a river system: subsequent differentiation events lead to increasingly specific trajectories, and increasingly specialized (and constrained) developmental trajectories. The other component is environmental acquisition itself, which we can approximate using Gibsonian Information. GI mediates the output of each sensory array, which serves to process and integrate direct sensory information. In combination with the canalization function, a subset of values dominate the output at different points in time. This is generally applicable to developmental contingency, where sensory units in an array are acquired at different points in time.

### 2.2 Architecture for Developmental Innateness

In developmental neurosimulation, an innate capability is the substrate of a morphogenetic or behavioral acquisition encoded in an artificial chromosome. An artificial chromosome is a vector that contains a genomic encoding, consisting of instructions that define the innate response and agent architecture. Our genomic encoding is represented on a binary string which represents the structure and function of traits related to the body (morphology) and innate behavioral primitives. While the mapping from encoding to substrate can vary, it typically follows a correspondence principle: features of the substrate are represented by a certain number of bits. In a dBV, we can use the morphogenetic substrate of light sensors as an example. A simple suite of light sensors have the following features: spatial coordinates for feature placement, multiple subsampling cells, aperture width of the sensor, angle of acquisition, and maintenance cells that provide real-time feedback. The entirety of the light sensor morphogenetic substrate occupies a certain number of bits in the encoding, while multiple sensors can result from duplication of these bits. The substrate of specific behavioral acquisitions requires additional bits and a separate encoding for features such as information bandwidth and visual angle. All parts of the encoding must be in place at the beginning of development, while the morphogenetic substrate must be in place to express the behavioral substrate (Figure 2). The corresponding encodings can be mutated or recombined in ways that modify function. Plasticity encodings allow for changes in encoding expression in real-time, often in response to environmental information (see Figure 2).

The origins of innateness in our computational agents is implemented through evolutionary computing techniques. One example of an architecture comes from the BraGenBrain platform [7]. Briefly, features of a phenotype are encoded on a linear artificial chromosome. These features are directly related to the size of the agent's neuronal network. A population of agents is generated, and each agent is subject to recombination and mutation. Over time, this population converges to a single set of navigation behaviors. While the BraGenBrain model works in implementation, the developmental aspects of this model is limited: for example, there is no canalization nor expression of artificial chromosomal encodings. Extending the BraGenBrain encoding strategy to a general architecture for dBVs (as well as other instances of developmental neurosimulation), we can refine this model in a number of important ways. Generally, dBVs are clonal (self-reproduction) that are subject to recombination and mutation. As these genomic encodings (contained on an artificial chromosome) and their associated phenotypes (the shape and transformational properties of an agent body) grow in complexity, it requires both the coordination of innate capabilities and the selective timing of innate capability expression.

*2.2.1 Coordination of Innate Capabilities.* Expression of the genomic encoding into multiple realizations of an agent is based on the biological observations of [43], in which cell behaviors and expressed genes are regulated through weak regulatory linkage. Weak regulatory linkage characterizes genetically-defined units linked by shared regulatory components. There tend to be multiple ways to activate a weak regulatory relationship, making our genomic encodings robust to a plastic response. Weak regulatory linkage is linked to the concept of evolvability, or the degree to which an organism or lineage is prone to evolutionary innovation, and affects embodiment by coordinating development of a phenotype through common function and location. When regulatory linkage is weak, traits that are linked together in a non-deterministic fashion. This allows sensors on the left and right side of the agent's phenotype to develop in parallel, or other symmetries to be enforced. The sensors not only become functional in an embodied context, but also experience a common critical period that allows them to synchronize their acquisition of sensory information. A binary encoding (shown in Figure 3B) can be mapped to newly-born neurons in the vehicle's internal model (Figure 3A), and experience variable expression due to experience over developmental time. These artificial chromosome encodings have two types of switches: identity and regulatory. Each one maps to a particular gene order (m) and level (n) combination.

This arrangement of cell identities in the binary encoding mimics the arrangement of Hox genes, which maintain body segmentation and anatomical order in animal development. Rearranging the gene order can provide compositional variety and novel wiring arrangements. Artificial chromosome encodings can also serve as instructions that modify the arrangement of phenotypic components. Timing of critical periods can be adjusted across a population or between populations according to the expression of the genomic encodings. Components of the agent phenotype can be continually reconnected according to their activation strength across different stages of learning (see Figure 3C). In this example, regulatory elements are activated to trigger the birth of components (neuronal network cells or body components), then express an activation strength fluctuating according to an encoding-specific distribution. Once a cell is wired to the sensor, however, the environment can modulate this activation strength. The internal network's collective behavior determines the canalization function and specialization of a vehicle's behavior. Innate capabilities serve to guide behavioral training. Adaptive behaviors and behavioral variety results as the neuronal network changes in time.

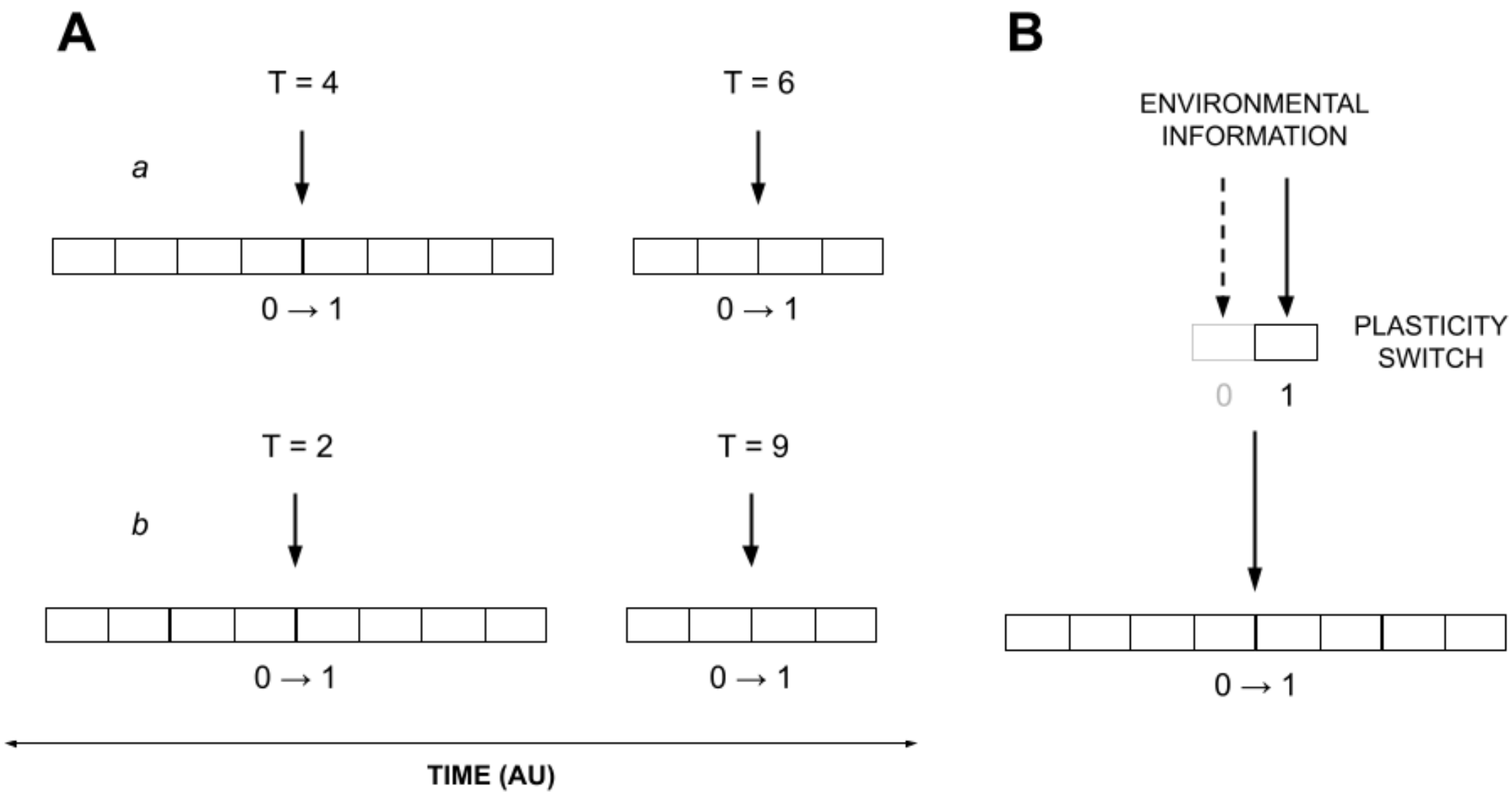

Figure 2. An example of an encoding via artificial chromosome (A), and an alternate form called the plasticity encoding (B). A shows two scenarios for the morphogenetic and behavioral substrates. *a* shows the expression of these traits at sequential points in time, and *b* shows changes in timing both earlier and later than the initial condition.

This arrangement of cell identities in the binary encoding mimics the arrangement of *Hox* genes, which maintain body segmentation and anatomical order in animal development. Rearrangement of this ordering can in theory provide developmental variety and novel wiring arrangements, and this architecture is extensible to incorporating additional artificial chromosome encodings as instructions that modify both the layers and arrangement of cells. The distribution of birth times and timing of the critical period (fluctuation in birth times and activation strengths) of individual cells can be adjusted across a population or between populations according to the expression of the artificial chromosome encodings. These cells can be continually reconnected according to their activation strength across different stages of learning.

A demonstration of how cells are born, die, and activated over developmental time is demonstrated using pseudo-data in Figure 3C. Briefly, a regulatory element is activated to trigger cell birth (a non-zero activation strength), then expresses an activation strength that fluctuates according to a distribution determined by the artificial chromosome encodings. Once a cell is wired to the sensor, however, the environment and other cells (acting through the sensorimotor signal) can modulate this activation strength. The collective behavior of this internal network determines the state of the canalization function, and with it the specialization of a vehicle's behavior in developmental state space. Innate capabilities serve as a substrate for behavioral training, and can be evolved and selected upon using a connectivity matrix of activation strengths for each cell.

*2.2.2 Heterochrony and the expression of encodings.* Encodings are expressed at specific times in the developmental process, and can be optimized for specific computational agents. However, the timing and sequence of expression can be moved forward or backward in processes that resemble dynamical and sequence heterochrony, respectively [44]. Heterochrony is a genetically-determined mechanism by which organisms (or in this case, agents) experience relative changes in the timing of component differentiation, rate of growth, or the general duration of development. For example, changes in the position of effectors relative to body length results in consequences relative to motor control. As there tends to be variation across organisms, heterochrony can provide differences in the size and shape of the mature dBV. Heterochrony is a Figure 2A shows the role of sequence and dynamical heterochrony in terms of both the ordering of morphogenetic and behavioral substrate encodings on a virtual chromosome, and changes in timing of expression of these traits in the agent phenotype ($T$ = time of appearance, arbitrary time unit). While sequence heterochrony (the ordering of developmental events) is important for maintaining a viable mature form [45], dynamical heterochrony is also important for extending the growth trajectory or limiting the size of certain features. In Figure 2A*a*, morphogenetic and behavioral traits are expressed closer in time, while in Figure 2A*b*, the morphogenetic substrate is expressed earlier in developmental time and the behavioral substrate is expressed later in time.

## 2.3 Agent Embryogenesis

One key aspect of continual developmental neurosimulation is the development of agent phenotypes from innate capabilities for learning and behavior. The source of innate capabilities are not only provided through artificial chromosome encodings, but in singular, undifferentiated precursor phenotypes that both reproduce and exhibit morphogenesis. Agent phenotypes (bodies, sensors, and internal networks) emerge according to a series of transformations shown in Figure 4. The transformations proceed from a single undifferentiated cell, transitioning to a three-layered embryonic form, which then results in a mature phenotype. These tissue layers (hereinafter referred to as types) are determined by mature function, and are ordered by their relative size in the final adult form. For example, if more sensors or effectors are needed, the black type will be bigger and located as the central ring in a three-layered developmental phenotype.

The three developmental stages shown in Figure 4 results from a set of transformation conditions, encoded in our artificial chromosome encodings and applied to the developing phenotypic applied at each stage. The difference between stages 1 and 2 is defined by two differentiation events and a sorting of types by order of size. Creating initial differentiation events not only allows the agent to attain more complex phenotypes, but also raw materials for later transformations. Each blastic layer shown in Figure 4 becomes a type that is restricted to a range of potential phenotypic components. As per Figure 4, the light gray type is restricted to forming the outer part of the body. The dark gray type is restricted to forming neuronal network and actuation components (e.g. muscles and joints), and the black type is restricted to sensor and effectors. This allows for a balance between developmental specificity and reconfigurability.

### 2.4 Positional Information and Connectivity in dBVs

Our architecture for developmental innateness is focused on the specification of both the dBV phenotype and neuronal network. Positional information is a more general mechanism behind the Hox gene analogy for the arrangement of cell identities in the binary encoding discussed in section 2.2.1. Embryonic development of the agent body is in part defined by the positional information of each cell and tissue relative to the whole body. In a biological context, cells and tissues acquire positional (coordinate-based) information based on their location relative to coordinating signals [46]. This allows us to build a body and neuronal network without an explicit blueprint. As a result, these positions become part of an addressable system of interactions which give rise to spatial segregation, movements of components from the same blastic layer, and global pattern formation.

In Figure 3, transformation conditions are accompanied by movements of layer components rather than cells or tissues. The black layer divides and extrudes to form a single frontal sensor and two rear side wheels. This requires positional information based on functional requirements and the relative position of the type in terms of germ layer organization. Positional information can also be characterized in terms of bits. Coordinating signals can be summarized as discrete molecular states that in biological systems differ between cells of a distinct tissue type [47], but in our case are specified by innate factors. Differentiation into body parts represents a gain in information content. The addressable cell location can also be represented as information content (bits), where a location is embedded in a local concentration gradient of patterning signals [48]. This can be represented as a stochastic gradient, which when combined with considering how events unfold in developmental stages (Figure 4) opens the door for applying various gradient descent models [49] to our model of dBV phenotypes.

Neuronal network growth and plasticity proceeds through connectivity-activation encoding, in which the matrix $W(i,j)$ of potential connections between internal nodes is shaped in a selective manner over developmental time. Expansion of the matrix represents an increase of cell number in the network, and the rate of expansion from the input-output initialization can either be deterministic or stochastic. All non-zero values represent active connections between all cells existing at a given time step. Updating $W(i,j)$ during the morphogenetic growth phase of development is governed by a discrete genetic algorithm and encoding featuring recombination, mutation, and selection (for further technical details about BraGenBrain, see [7]).

Regardless of the exact method of generating variation in cell number expansion and active connections, we still must evaluate a large number of candidate network topologies. Thus, our genetic algorithm utilizes a fitness function that is determined by movement of the vehicle relative to target stimuli. The fitness function evaluating a given dBV's *W(i,j)* matrix is based on stimulus-to-vehicle distance and force generated by vehicles in response to a given stimulus. Even though the introduction of variety randomizes matrix elements over time, selection nevertheless favors the emergence of networks that enable recognizable behaviors [8].

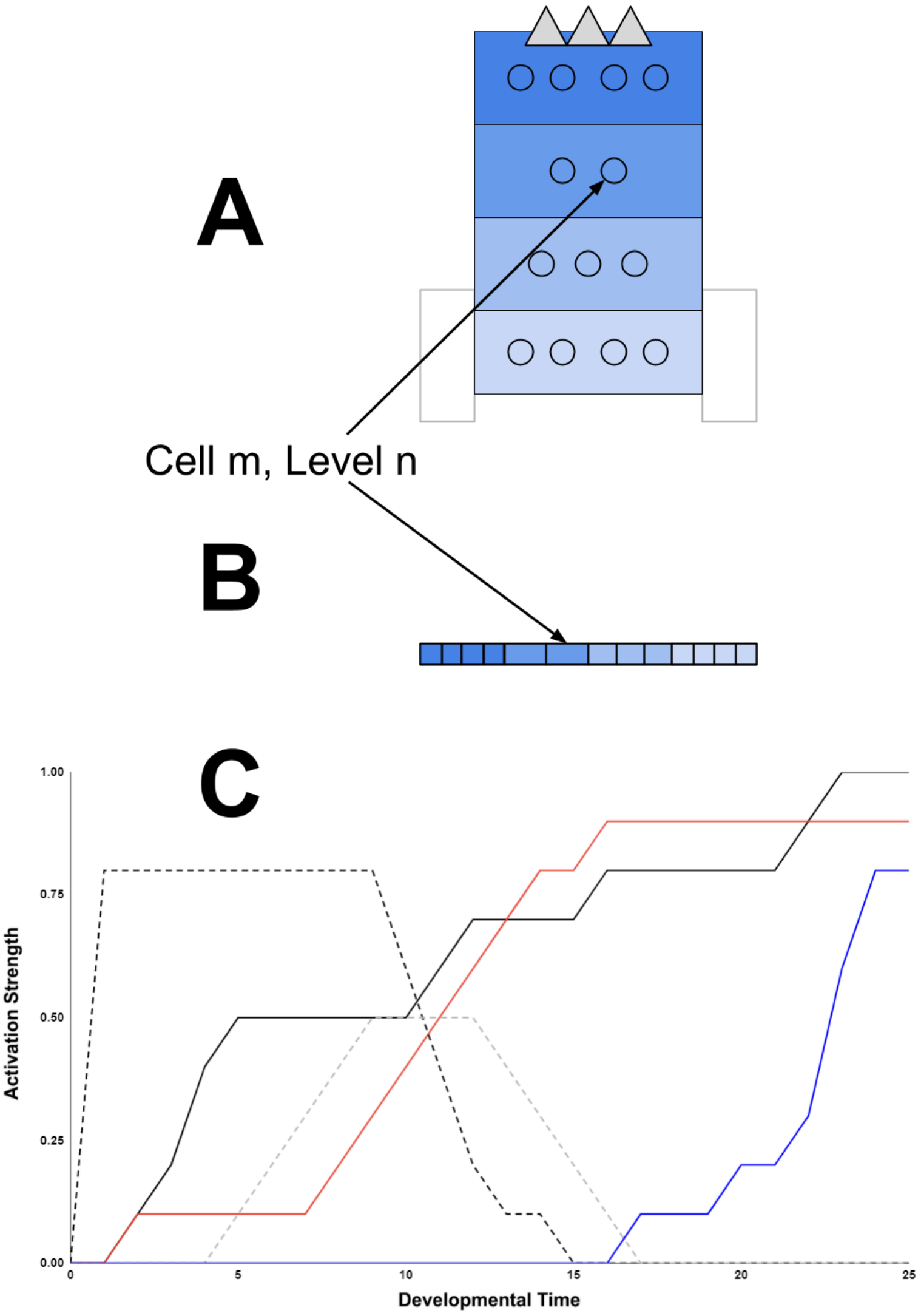

Figure 3. Demonstration of how morphogenesis and developmental time proceed in a dBV. A: a dBV with four levels of internal nodes (interneurons between the sensor and effect). B: a linear binary encoding that maps to each cell and level combination. C: the behavior of five neurons as they are born (or die) and acquire (or lose) an activation strength over developmental time. Black and gray dotted lines are cells that are born and then die off during development. Black, red, and blue solid lines show birth and increasingly strong activation over developmental time.

### 2.5 Mechanisms of developmental plasticity and learning

Using a genetic algorithm as a training mechanism also generates a variety of topological orderings of matrix *W(i,j)* through mutation and recombination. For larger dBV neuronal networks, an encoding of coordinated projection identity can be used to more closely mimic a biological nervous system [50]. In the scheme introduced here, the critical period corresponds to a period of accelerated evolution in which the mutation rate is increased. This increases the number of potential configurations, but using a fitness criterion ensures that all of them are capable of learning. At the end of the critical period, the topological mutation rate is set at near zero (or turned off completely). In previous work, it has been demonstrated that associative learning can occur on a stable (post-critical period) *W(i,j)* matrix [7, 10] using a Generalized Hebbian Algorithm (GHA). Associative learning in dBVs is instantiated by building associations between sources of multiple senses simultaneously, and is thus the essence of developmental freedom. In [7, 10], olfactory and gustatory stimuli are distributed independently in a multidimensional space. The associative learning process acts to change the connection weights in matrix W(i,j) according to the co-occurrence of the two stimuli. In this way, spatially-explicit multimodal associations can be acquired.

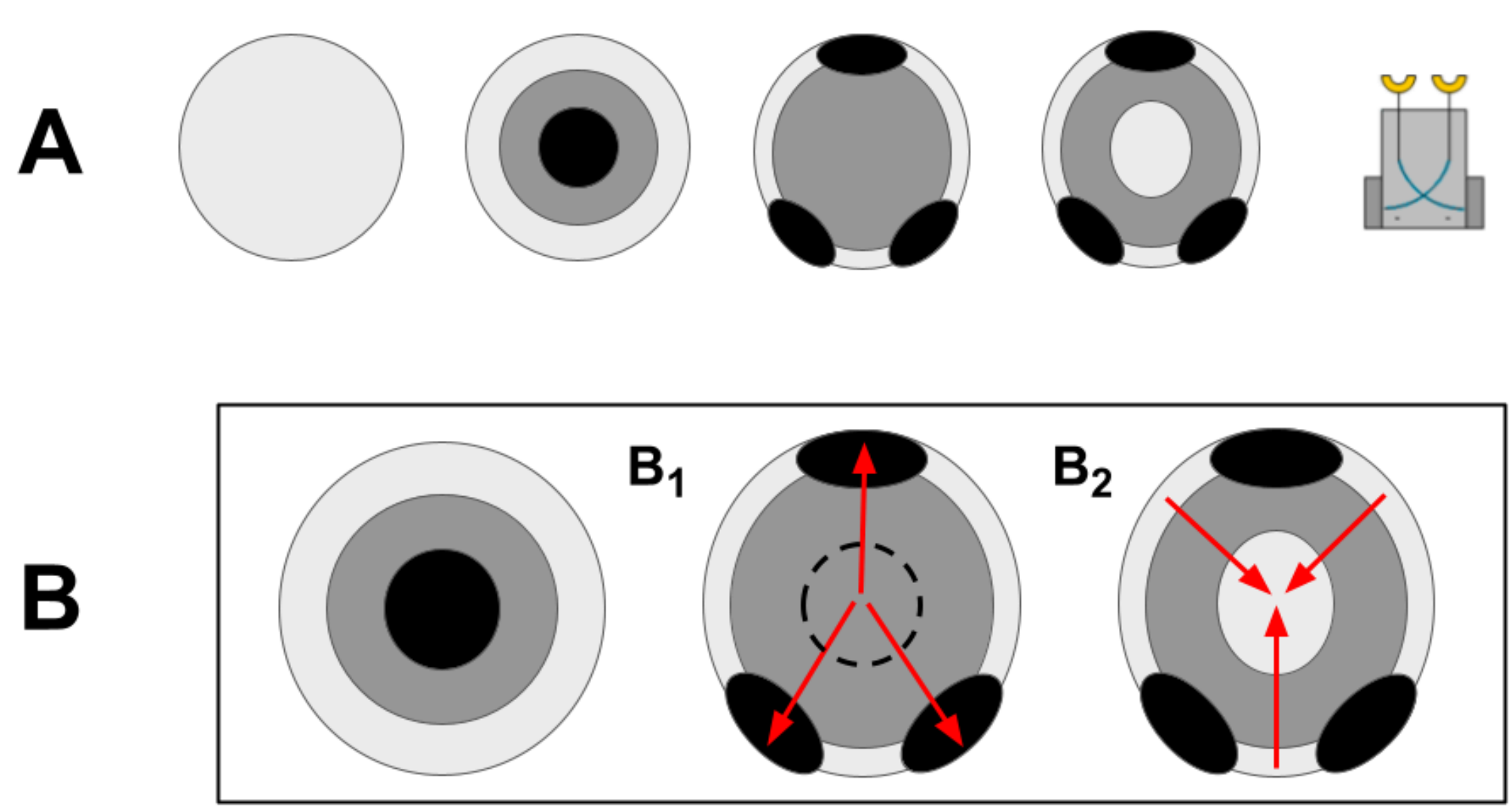

Figure 4. The embryonic development of a dBV from a single phenotype to a mature vehicle. A: progression of embryonic development from an undifferentiated type to a three type phenotype, to a phenotype with proto-sensors/effectors, to a phenotype with a proto-connectome, to a mature vehicle. B: the three type phenotype, phenotype with proto-sensors/effectors ($B_1$), and phenotype with a proto-connectome ($B_2$) with red arrows showing the migration of types to other regions of the phenotype. In $B_1$, the inner type (black) migrates to the edge of the phenotype, forming one proto-sensor and two proto-effectors. $B_2$ shows the formation of a proto-connectome as partial migration of the outer phenotype (white) inward to the former location of the inner type (black).

## 3 Broader Methodological Concepts

### 3.1 Gibsonian Information and Naturalistic Behavior

dBVs experience sensory flow as they move around their environment with respect to environmental stimuli. The dynamic nature of the reference frame itself involves active direct perception [51] and can be detected using an approach called Gibsonian Information [41]. The latter might be identified using quantitative measures that produce lawlike relationships [52]. Aside from taking advantage of changes in timing, our developmental system also exploits relative information, or the continuous difference between multiple sensors. The current sensory state allows for the detection of coherent motion and spatial differences in a complex scene more generally. Angular differences between sensors in different locations across the front and sides of the vehicle capture the intensity or shape of a stimulus. dBV receives continuous sensory input from this stimulus. One way is to integrate the information using the GHA, while another way is to compare incoming sensory information to movement of the vehicle. This latter operation is a recurrent mechanism, and can be fully utilized by implementing other types of learning algorithms.

### 3.2 Differentiation Trees and the Evolution of dBV Lineages

Aside from biological and cultural evolution, we can also use phylogenetic techniques to understand the developmental origins of specific morphogenetic acquisitions. We use a differentiation tree-inspired model [53, 54] to represent the potential for developmental variation amongst dBVs. In biology, differentiation trees provide a means to understand waves of expansion and contraction that contribute to the emergence of new tissues in the embryo (Figure 5). For dBV agents, expansion and contraction waves represent phenotypic expansion (gain) and phenotypic contraction (loss) that unfold to the right and to the left, respectively. This provides us with basic conditions for growth and morphological changes that translate into different conditions for embodiment.

A differentiation tree represents all possible developmental pathways through which a population of dBVs can proceed. At each branching point, a transformation may occur, leading to developmental phenotype contingent upon its past developmental trajectory. Each colored portion of the differentiation tree in Figure 5 begins and ends at a differentiation event. For example, the red branches of the differentiation tree define a differentiation event (originating at a blue node) that transforms all possible configurations of the phenotype until the next viable differentiation event (shown in green). In this example, there is an non-viable differentiation event marked in red, which leads to a transformation that can make the dBV non-functional (the red branch in the middle of the tree leading to no differentiation event). This represents a transformation in the phenotype. These transformations can have an effect on future neuronal network development, particularly if the transformations have an effect on either sensors or effectors. For the purely developmental example in Figure 5, contraction waves represent restricting the phenotype and results in a lower growth rate. A version of the differentiation tree adapted to developmental neurosimulation is shown in Figure 6, with the branching process proceeding downwards instead of upwards. In Figure 6, this translates into a loss of agent parts, representing progressive degeneracy the farther left in the tree we move. By contrast, expansion waves in Figure 5 yield an expansion of the phenotype and a higher growth rate. Figure 6 adapts this to a gain in phenotypic parts and an overall increase in phenotypic complexity.

Building a differentiation tree for agent development is useful for composing BV phenotypes in development. In agents, contraction and expansion waves represent organizational rearrangements and growth of the ancestral phenotype. In Figure 6, we can see the emergence of several dBV lineages representing variations on basic innovations. A single branch of the differentiation tree results in two vehicle morphologies. Some of these are viable, and some are not. In the Figure 6 example, we first see axial expansion of square body elements, followed by the emergence of triangular motifs and various rearrangements and ultimately proliferations of the basic dBV phenotype.

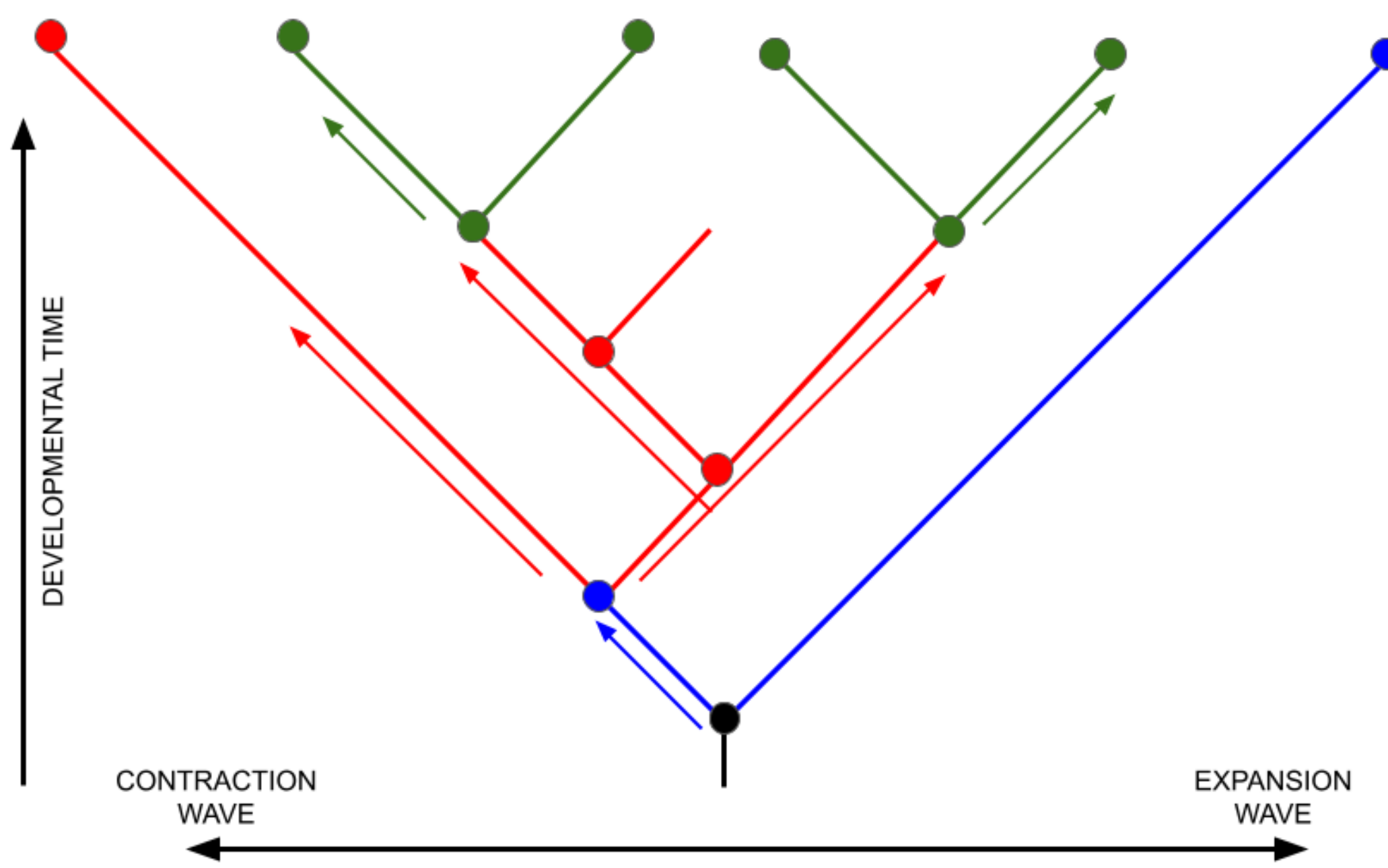

Figure 5. A differentiation tree: a branching model that represents the differentiation of tissue types in the embryo. This example will map to the dBV context with modifications as necessary. Each color (black, blue, red, green) is a differentiation event that proceeds from another (black → blue → red → green). These events result in one contraction wave (produced restricted phenotypic modules with a lower growth rate) and one expansion wave (produces expanded tissues with a higher growth rate).

### 3.3 Potential for Reinforcement Learning

In general, dBVs are a special case of a reinforcement learning model. Yet there are differences between developmental and reinforcement learning. The first major difference is that feedback through reinforcement is not made explicit during the developmental process. While a dBV receives environmental feedback, this is often decoupled from the expression of an internal set of processes. The dual process of generating a neuronal network and learning on the network could be formulated as a reinforcement learning problem. We argue that this constant interaction with the environment scaled with time plays a significant role in the development process of the dBVs. Our dBV model can be contrasted with the dual world/self model of [55]. In our case, we can benefit from reinforcement learning through disentangling the intrinsic and learned representations which contribute to the developmental process of adult behaviors.

Rather than the Genetic Algorithm/GHA implementation, implementation of a model-free Reinforcement Learning paradigm is possible, where the continuous sensorimotor space grows temporally; this approach is more continuous across developmental time and work while the agent is learning a policy. In such a case, learning could be optimized as a function of morphogenetic growth. This might improve performance in the adult phenotype. As for the effects of critical period timing, maximizing the consequences of sensory feedback during this time would allow us to take advantage of the correspondingly intense rate of morphogenetic growth.

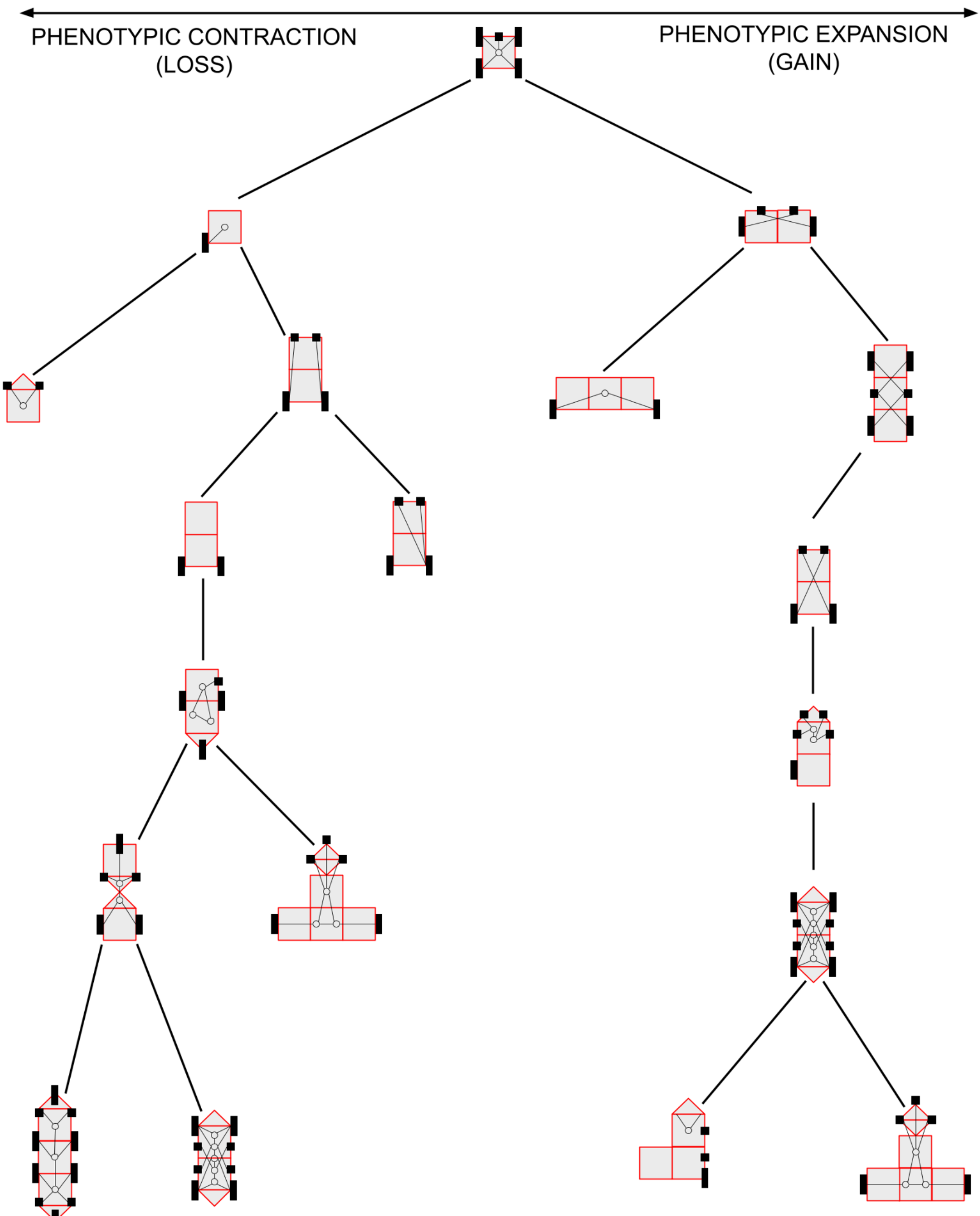

Figure 6. Developmental composition of BVs using a binary differentiation tree.

Developmental Braitenberg vehicles are models of embodied nervous systems that navigate and constantly interact with their environment. To implement an RL model of dBV learning, we can appeal to the principle of parallel, loosely coupled processes. This principle states that intelligence emerges from the agents' interactions with the environment. These interactions are based on loosely coupled processes run in parallel and connected directly to the sensory-motor apparatus of the agent [56]. However this continuous sensorimotor space which is inflated with time along with the large continuous action spaces makes it difficult to learn good strategies (train optimal policies) to perform a task, some approaches have been proposed to counter this problem- for instance, adding developmental layers to the neuronal networks this could help the agent to learn better policies, embedding information about the source task in the reward function [57] and also applying other curriculum learning [58] techniques which we believe requires further consideration.

# 4 Analysis

## 4.1 Contingency Analysis

Aside from being characterized by generativity and plasticity, development is also characterized by developmental contingency [59, 60]. Consider a neuronal network resulting from an innately-defined relationship between sensors and effectors. Each time step introduces components such as an intermediate node or the formation of a new connection between two nodes. At the end of development, a complex nervous system results. We hypothesize that a neuronal network where no circuit components can be removed without functional consequences results in a representational capacity quite unlike a traditional BV. From the initial input-output mapping, every component generated by the developmental process essentially locks the network into a subset of possible future topologies. Topological evolution is an internal embodiment of path dependency. From a design perspective, our choice of critical period timing has an effect on what the vehicle can learn and how it behaves.

## 4.2 Different Scenarios for the Critical Period and Developmental Freedom

Figure 7 demonstrates how dBVs respond to changes in the timing of the critical period and the resulting onset of developmental freedom. Consider two scenarios: early onset of the critical period and late onset of the critical period. The evaluation of early and late onset is defined by when the addition of nervous system elements occurs (early or late). Early critical period onset means that relatively few nervous system connections are generated after the end of the morphogenetic period, occurring mostly within this period. By contrast, late onset allows for a larger proportion of nervous system connections to be added after termination of the morphogenetic phase. Early and late developmental freedom are a direct consequence of critical period plasticity timing, but the timing of this process also has unique signatures on the adult neuronal network.

*4.2.1 Early Critical Period results in Early Onset of Developmental Freedom*. When the critical period occurs early, the adult neuronal network tends to be more homogeneous, exhibiting the same types of configurations and motifs. In this scenario, there is less time for preliminary or auxiliary nervous subsystems or behavioral traits to develop. There are also fewer varieties of experience prior to maturity, but since the neural topology is more constrained, the capacity for developmental freedom is homogeneous across individual dBVs. The experience of an early critical period results in a high degree of overall developmental freedom. An early critical period allows for faster ascendance to a fully mature (adult) suite of behaviors suite and capabilities, and thus more exploration of network weights on a stable topology. This enables more refined behaviors and associations to emerge and for shared experiences between individual dBVs to be more common.

*4.2.2 Late Critical Period results in Late Onset of Developmental Freedom*. When the critical period occurs later, the adult neuronal network can exhibit a larger number of possible configurations across a population of dBVs. Delaying the onset of the critical period affords more time for exploration. This results in a slower ascendance to the mature behavioral suite, and provides the adult phenotype with a greater number of alternate network topologies, which can lead to divergent capacity for developmental freedom across individual dBVs. By contrast, dBVs that experience a later critical period results in a low degree of overall developmental freedom. Due to a later critical period, there is a slower ascendance to adult behaviors and capabilities. The consequence of this is less exploration of weights on a stable network topology, which results in more stereotyped behaviors and associations. However, due to divergent network topologies, we should see a heterogeneity of function, and less shared experience when comparing individual dBVs.

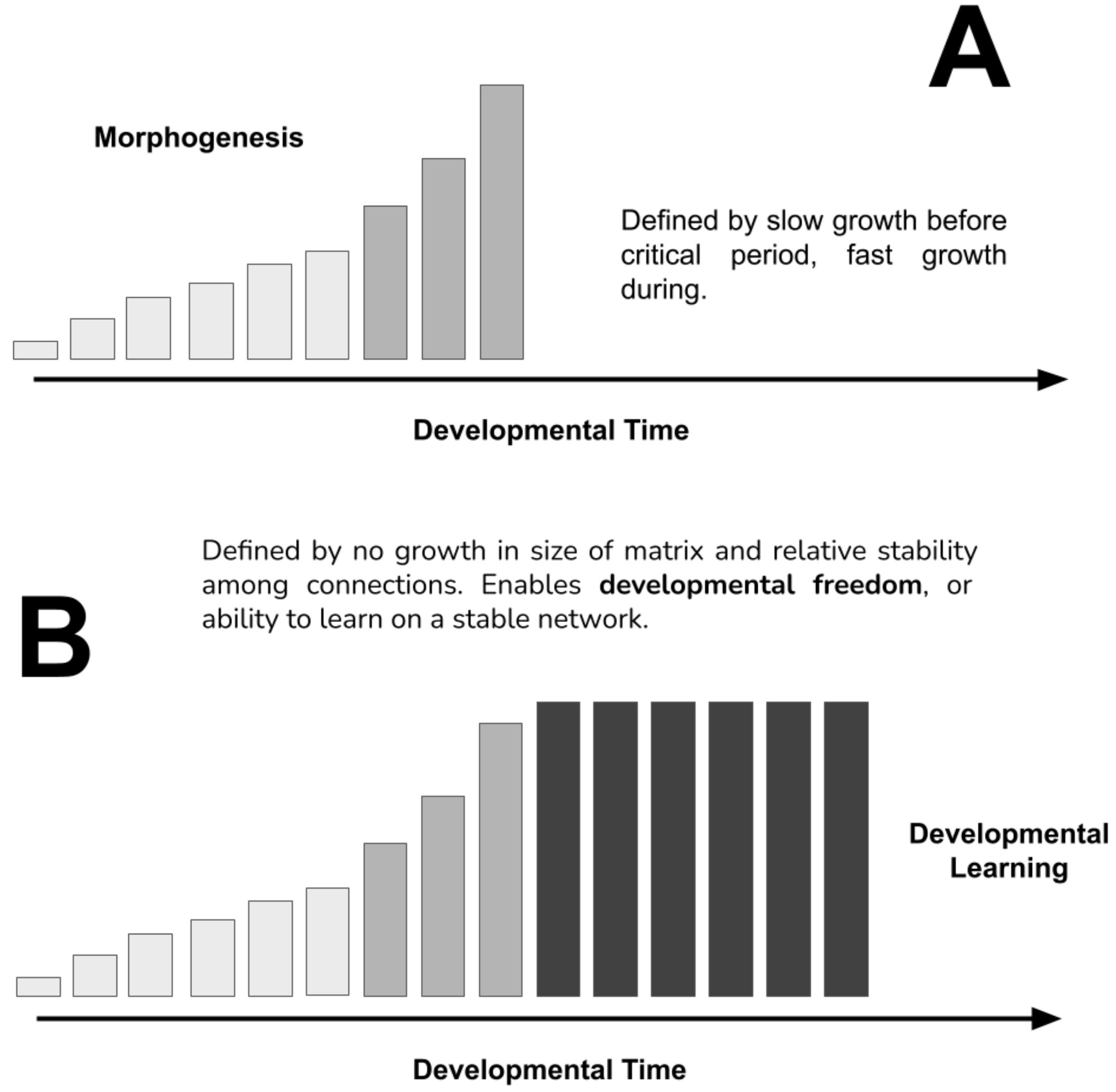

Figure 7. Diagram showing the relationship between the morphogenetic period (light and dark gray), the critical period (dark gray), and developmental learning period (black). Developmental freedom is defined by the black squares, while the shifts in the critical period that define our hypotheticals occur along the axis of developmental time.

Our conception of an agent neuronal network goes beyond what is typically found in a Braitenberg Vehicle in two ways: continual fluctuation in the number of internal processing units and connections, and formation of recursive structures such as internal processing loops and sensorimotor feedback. This forces the system to adapt through the adoption of adaptive capacity rather than specializing towards specific optimal outcomes [61]. In dBVs, morphogenetic events result in two types of feedforward elements: singular connections from A to B, and composite connections from A and C to B (or from A to B and C). While the singular connection type provides a linear feedthrough of the sensorimotor signal, composite connections provide nonlinearities such as convolution, synthetic effects such as cross-talk, and even the potential for interference that can degrade the sensorimotor signal.

### 4.3 Canalization Function

We propose that the Genetic Algorithmic and Hebbian Learning models can be unified using a developmental loss function based on a canalization function [62, 63]. A canalization function is a binary switch that when triggered by the canalyzing selection of a non-uniform environment [64], can shift the developmental state to a new trajectory [65]. This is useful for modeling branching processes that describe restricted regions of developmental possibility space for specific neuronal network topologies [66]. In this case, the canalization function enforces branching after a certain number of morphogenetic events despite slightly different explorations of developmental possibility space [67]. The alternate morphologies featured in Figure 6 can lead to developmentally non-viable vehicles, or present novel means to process a vehicle's sensorimotor signal.

### 4.4 Agent Developmental Cognition at Different Scales

Figure 8 demonstrates how continual learning is not only a property of a single time scale (e.g. encoding and recall of a remembered stimulus), but a property of the entire lifespan as well as intergenerational cultural and biological inheritance. In Figure 8A, we see a timeline of different cognitive events alongside their time course. This comparison is to place different types of events in context. We expect that agents will not need to experience these events directly in order to experience them. Rather, they develop the conditions suitable to experiencing these events.

Figure 8B shows the same timescale, but with different types of behaviors manifest at different durations of time. When these behaviors occur at different time scales (e.g. single stimulus presentation, lifespan), they are said to be compositional: long-term memory consolidation builds upon neuronal activity, as historical events build upon short coffee breaks. The precursors of compositionality are enabled by a canalization function that implements development of the phenotype and neuronal network in timed stages. In this way, learning is structured by multiple developmentally-inspired mechanisms. This is particularly useful in the phenomenological sense [68], particularly in terms of producing an agent's response at multiple time scales. In connecting phenomenology with multiple timescales, a hierarchical model has been proposed by [69].

*4.4.1 Timescales of development and learning.* Integrating embodied agents with a developmentally bio-inspired approach to neurosimulation provides a unique opportunity to study these multiple timescales. In a population of agents, we observe some agents in egg form, others in developmental mode, and still others in reproductive mode. This provides multiple life stages and generations against a background of environmental information and cultural practices. At some life stages, this background can serve as both directly and indirectly influential to learning. For example, neuronal activity and fast recognition is important for training during expression of innate capabilities (morphogenetic and behavioral substrate). Experiencing a coffee break and consolidating short-term memory is critical for building on top of the agent neuronal network. Historical events are relevant to intergenerational learning and strategies for reproduction. Replicating the complexities of organismal biology and its behavioral milieu will improve our chances of simulating intelligent behaviors far beyond task-constrained and human-centric examples.

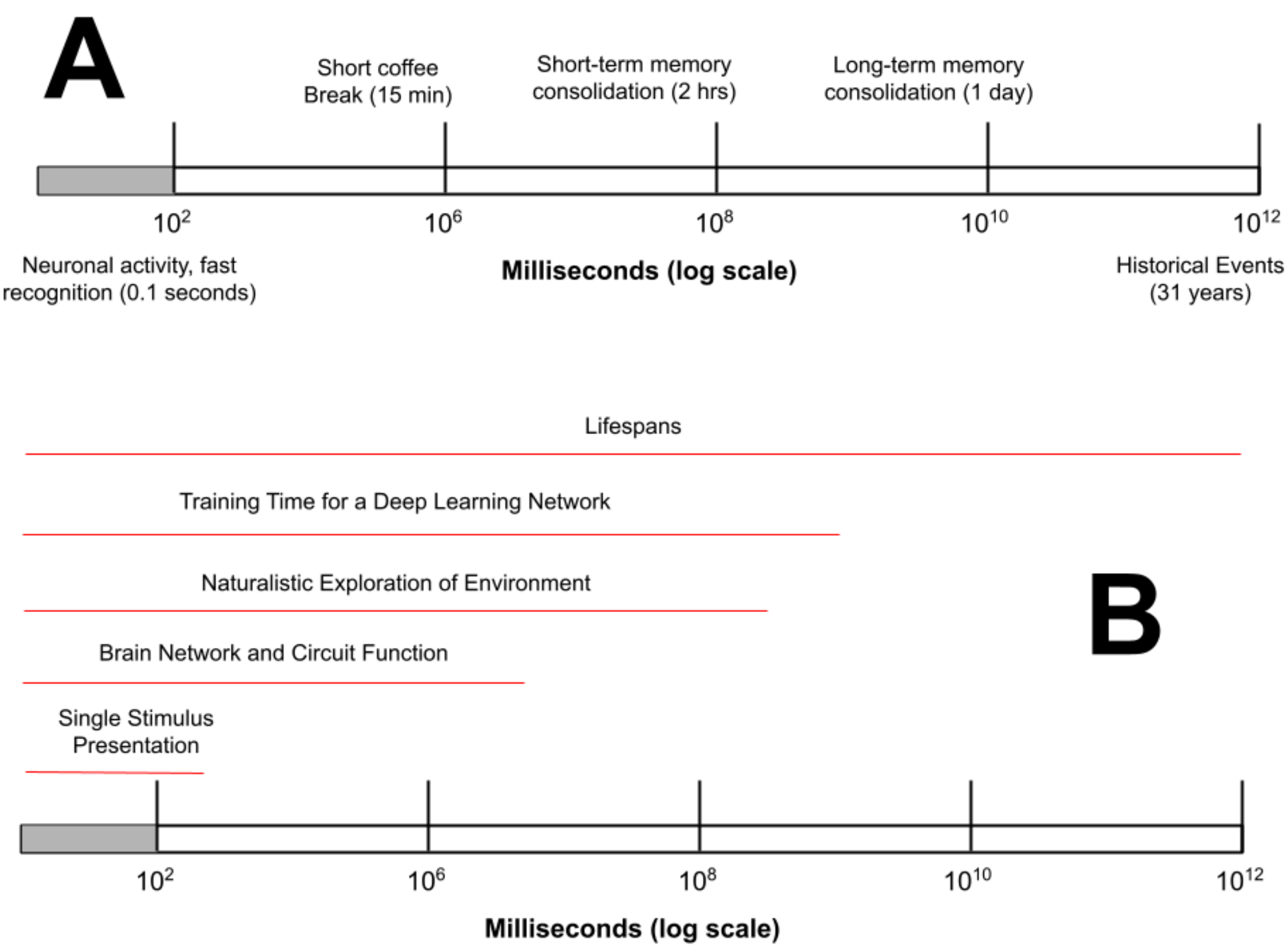

Figure 8. A labeled timeline of agent development as a continual process, from short timescales of cognition to multi-generational inheritance and interactions. A: a log-millisecond (ms) timeline of naturalistic cognition with various events and their approximate duration (approximated in seconds, minutes, hours, days, and years). B: a log-millisecond (ms) timeline with different aspects of agent biology, artificial neural networks, and interaction with the environment and their approximate duration.

*4.4.2 Meta-brain dBV modeling.* A meta-brained dBV can be used to model the multiple modes of agent cognition as shown in Figure 9. The meta-brain demonstrates two modes of cognition: the low-representation state is indicative of smaller timescales of neural activity and pairwise connectivity, while the high-representation state is indicative of longer timescales at which event-related phenomena are integrated. One way to conceptualize this is in the form of neuronal avalanches [70, 71]. Neuronal avalanches are large-scale responses that result from the accumulation of small changes. The sandpile model [72] describes this as large-scale displacements of sand in response to the stochastic grain-wise growth of a sandpile. Sandpile displacements are the product of nonlinear responses to additive growth. While this has been demonstrated to be a property of neuronal dynamics, it is also much more complex than an input-output (sensor-effector) relationship. In the agent's neuronal network, events that respond from short-timescale activity accumulate and get passed to the high-representation state. In this layer, accumulation of activity can yield large-scale changes in the output behavior [73]. For example, a 15-minute interval of naturalistic behavior leads not to a linear repetition of simple outputs, but multiple transitions and multiplicative combinations of output behavior.

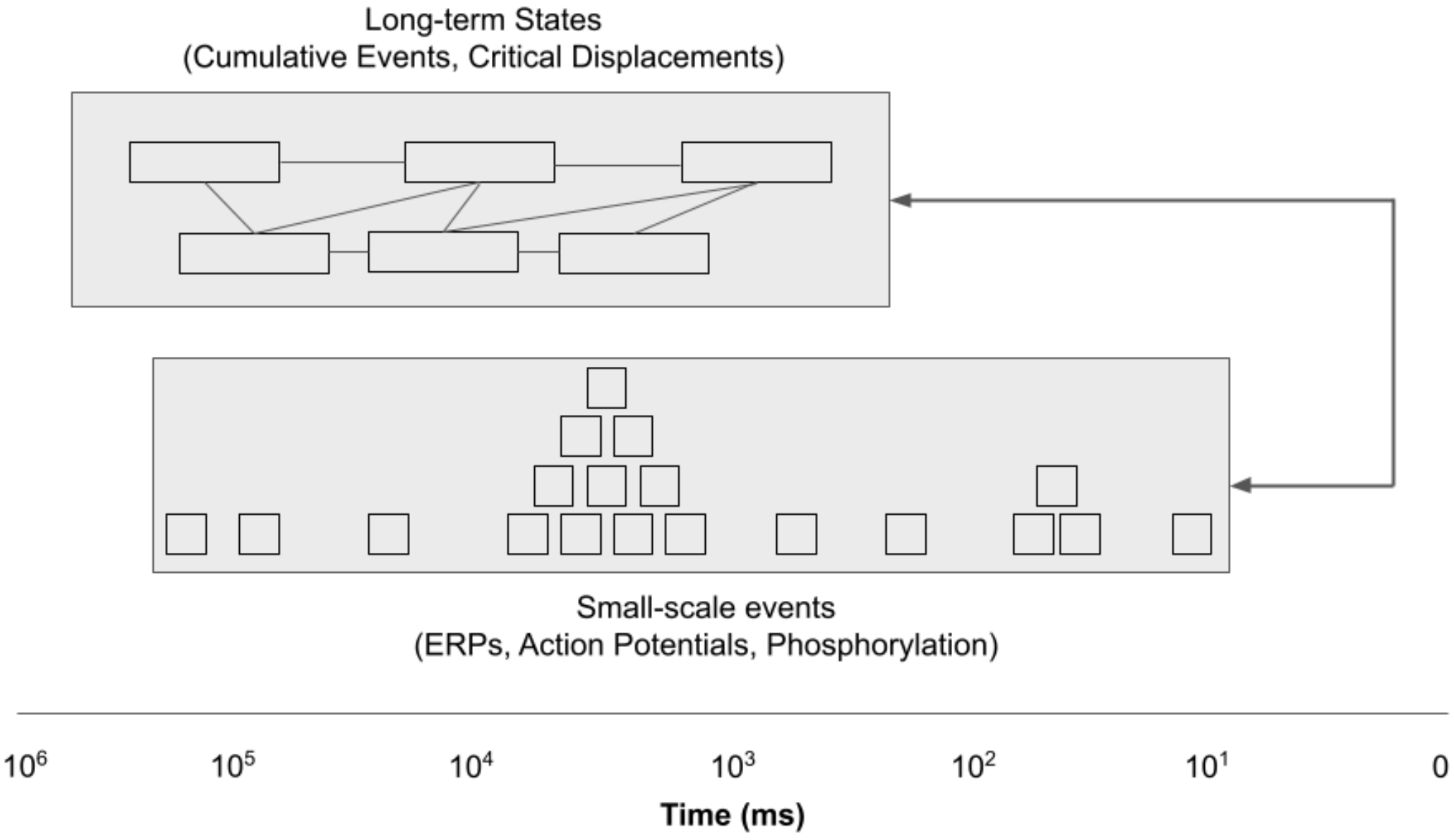

Figure 9. Subcognitive networks (corresponding to small-scale events) and cognitive networks (corresponding to large-scale events) modeled using a meta-brains configuration. A log-millisecond (ms) timeline of naturalistic cognition with various events and their approximate duration is used to characterize their place in time.

*4.4.3 Development of layers as enabling global response.* The advantage of the meta-brain model is to demonstrate how small-scale events in individual neuronal network units can be translated into global states. Meta-brains allow for these incompatible descriptions to be summarized by leveraging the different representational complexity of laminar organization. Laminar complexity allows for compartmentalization of information as it is processed internally: single units can serve as localized and specialized processors, while brain-wide states that interact with the surrounding somatic layers can also play a role in information processing. This can also provide a link between more complex behaviors. For example, steering behaviors that respond to 3-D curvature can be approximated using taxis in traditional Braitenberg Vehicles [74]. This can be modeled with representations using non-linear control systems or ODEs, but this specifies reactive behaviors without making the connection between neural microcircuits and global states. Walter et.al [75] propose that neural microcircuits are fundamental units of processing in artificial neuronal networks, and replicate these in artificial systems using heterogeneous network topologies. In addition, we can use multi-layered representations to model social behaviors, which involve both neurobiological mechanisms and behavioral interactions [76]. As an enabler of developmental generativity, the subcognitive/global state meta-brain improves upon Braitenberg's original typology of simple behavioral states [8].

## 5 Discussion

### 5.1 Broad Overview

In this paper, we have presented a broad overview of the developmental neurosimulation approach to embodied cognition, focusing on developmental Braitenberg vehicles (dBVs). Taking an embodied developmental approach to agent simulation is superior to both canonical approaches to embodied intelligence as well as simply recapitulating developmental psychology. Approaches that simply account for development as a growing neuronal network is similarly insufficient. A sufficient developmental approach takes into account what Lux et.al [77] and Lickliter [78] summarize as embodiment and embeddedness. In both cases, embodiment refers to advances beyond the psychology of the agent [79]. As we argue with respect to phenomenology, understanding the world involves multiple timescales and multiple mechanisms at each timescale. Establishing the dBV in space also involves both a body-centric reference frame and a goal-directed frame. However, that goal-directed frame is simply relevant to immediate behaviors, not necessarily its more ultimate origins [80, 81]. For example, the energetic imperatives or various developmental transformations of the agent might conflict with the priorities of goal-directed behavior. This leads us to considering embeddedness as a factor in developmental neurosimulation. Agents of more than trivial complexity must deal with relations between physical, biological, and cultural phenomena. As these multifaceted objectives often do not point in the same direction [82], understanding the relational nature of these different contributors to the emergence of behaviors in development [83].

In [10], we have previously argued that dBV architectures are low-representation networks that require other models to compute context and other, more sophisticated representations. Yet it may be that aspects of the dBVs embodied nature may afford dBVs a more robust comprehension of the transactions made with their environment over developmental time [84]. We have also argued that developmental freedom, spatial embodiment, and critical period regulation play a role in learning that is missed using other types of artificial learning. The dBV approach may also be a means to balance the processing previously unseen information (unsupervised developmental learning) with the need for a set of innate capabilities. Aside from our previous example of canalizing selection, in which the environment is the perturbing factor, we can also use sensory deprivation to analyze the potential of dBV neuronal networks across development and in its mature state. Sensory deprivation allows for environmental inheritance to be disrupted in a way that affects propagation of the sensorimotor signal more generally. One example of sensory deprivation is the removal of sensors that provide a coherent sensorimotor signal. Blocking the sensory inputs on one side of the vehicle will lead to behaviors that are biased to the non-occluded side. Like morphological asymmetries, sensory deprivation does not affect the morphogenetic process, but does affect how sensorimotor signals are conducted. A more indirect example of sensory deprivation with respect to dBVs is a stray cat that does not have the proper socialization experience.

## 5.2 Reassessing Developmental Neurosimulation as Continual Learning

We can also put developmental neurosimulation in the context of continual learning methodology by thinking about dBVs as agents. The first methodological issue is *concept drift*. Concept drift is where various changes (ranging from gradual to sudden) in the agent's environment occur with respect to their representational capacity [85]. Statistically, this translates into how the distribution of events changes over time [86]. This is similar to the Sandpile model example shown in Figure 9. While the sandpile model allows for new conceptual frameworks to be attained in a non-linear manner, it too can also lead to potentially catastrophic shifts in function. Yet it might also be that naturalistic behavior can smooth out the transitions between concept classes by providing context in an embodied manner. Therefore, combining embodiment with a non-linear account of a non-stationary world might reduce the chance of catastrophic forgetting [87].

The other methodologies of interest are *class-incremental learning* and *forward transfer*. Class-incremental learning assumes that the agent's environment is non-stationary, and disallows for classes to be reused across experiential contexts [88]. In [89], a single context variable is used to compensate for shifts in accumulated knowledge. We believe that our approach tackles this same problem in multiple ways, including utilizing criticality to manage transitions and asymmetries in the neuronal network and morphology. Indeed, the incorporation of anatomical segregation [90] and detailed synaptic processing models [91] can resolve some drawbacks of non-stationarity. In forward learning, transferring what is learned in the past to tasks that are encountered in the future must overcome model loss characterized as catastrophic forgetting [92]. In both cases, plasticity and regeneration on a late developmental network can play a role consistent with these strategies.

## 5.3 Lessons from Embodied Cognitive Morphogenesis

There are a number of lessons from a related approach called Embodied Cognitive Morphogenesis [93], which is based on the developmental dynamics and emerging sensory systems of biological embryos. The first lesson is that computational agent development is governed by multiple overlapping networks. While Figure 1 demonstrates the emerging agent neuronal network, other networks such as the relationships between different artificial chromosome encodings, sensory structures, or morphogenetic differentiation overlap. Another lesson is that as an agent's neuronal network begins to grow in terms of internal processing units and connections, it becomes increasingly embedded with morphogenetic and behavioral traits that define the agent's phenotype. Thus, information processing becomes embedded in a more active spatiotemporal context. The advantage of developmental embodiment over static embodiment involves symmetry-breaking events that can be summarized by tree representations (canalization function and differentiation tree) at multiple scales. In our overlapping networks, symmetry-breaking also introduces a phenotypic information eddy (circular flows from laminar ones) that can enable further structure and feedback loops.

As these structural and functional networks come online, information processing is progressively embedded within a phenotype [94]. Importantly, symmetry-breaking events provide anatomical context for the emergence of cognition through the phenotype (body) and nervous system (connectogenesis). Figure 6 demonstrates how these developmental processes can result in phenotypic symmetry-breaking events, which in turn leads to differential sensory processing and potentially even generating behavioral states from small-scale phenotypic features. Both rooted (directed graphs) and unrooted (networks) tree structures are ubiquitous in the developmental neurosimulation approach. Embodied Cognitive Morphogenesis offers several additional formulations of tree structure derivatives, the basic lesson is that embodied cognition can be explicitly tied to the contingencies produced by specific developmental trajectories. Related to this are the transformational aspects of developmental phenotypes, which also intersect with the expansion of neuronal networks across development. In the case of morphogenetic substrates, different patterns of differentiation of computational agent germ layers into distinct features can potentially result in radically different neuronal network architectures.

## 5.4 Expressivity and Interpretability of Developmental Neurosimulation

As bio-inspired algorithmic systems, dBVs require not only high expressivity but also extensive interpretability. Expressivity is crucial for capturing naturalistic behaviors that unfold over continuous time at multiple scales. There is also a need to interpret the behaviors generated from these systems, which can help us understand behavioral complexity at multiple scales more generally. A few examples from dBV implementations have been identified for consideration.

The integration of multiple sensory channels in the dBV instantiation MultiLearn [7] can be implemented at either the periphery (among sensory features) or in the neuronal network. While interpretation of multisensory integration outputs is difficult, characterizing single integrator internal processing units and the multiplicativity of their interactions [95] provides a way forward. The collective behavior of dBVs must also be interpreted, and has been classified for the dBV instantiation BV collectives in [7]. Further systems of interpretation might involve self-observation of collective behaviors by individual agents.

In [7], the hinge model of potential relationships between genotype and behavior is quite relevant to understanding how changes in the artificial chromosome encodings within populations and across generations can be characterized. One benefit of having a formal model of innateness is that it allows for generativity with expressivity, contributing to the expression of coherent and relevant behaviors. While the structure of these encodings can be beneficial to studying these relationships, there is nevertheless much that can be learned from the interactions of innateness, embodiment, and goal-directed behaviors.

## 5.5 Summary

With our toy model of an emerging nervous system, we can show how embodied neuronal networks are more sensitive to heterogeneous data, particularly in the spatial context. Furthermore, dBVs also teach us about the emergence of behavior as being deeply intertwined with the developmental process [11]. We can see this in the zebrafish visual system, where the maturity of representations and functions go hand in hand with developmental changes in behavior [96]. Developmental Neurosimulation is also reliant on allostatic control and the strategic pace of connectogenesis. Further, the relationship between connectogenesis and developmental freedom as a means to enable learning is of particular interest for future work on related concepts. More generally, future work on the continual developmental neurosimulation approach will explore the potential of strategically shaping a neuronal network and agent body in tandem with multiple adaptive mechanisms. We also seek to understand the potential of using dBVs as a strategy for enabling complex unsupervised learning in an embodied agent. Ultimately, this provides a catalog of ways in which we can apply fundamental aspects of developmental bio-inspiration, particularly the lessons of Embodied Cognitive Morphogenesis, to the enterprise of Continual Developmental Neurosimulation.

## Acknowledgements

We would like to thank members of the Saturday Morning NeuroSim and DevoWorm research groups for their ideas and input. Particular thanks go to Brian McCorkle, Morgan Hough, and Amanda Nelson for their discussions and insightful comments, and Ankit Gupta for their work on the dBV instantiations. Thanks also go to the Google Summer of Code program for their financial support in helping to develop the developmental Braitenberg Vehicles (dBVs) approach.